\documentclass[12pt]{article}

\usepackage{axodraw}

\makeatletter

\def\paragraph{\@startsection{paragraph}{4}{\z@}{+2.00ex plus
 +1ex minus +.2ex}{1.5ex plus .2ex}{\it\normalsize}}

\def\section{\@startsection {section}{1}{\z@}{+3.0ex plus +1ex minus
  +.2ex}{2.3ex plus .2ex}{\normalsize\bf\boldmath}}
\def\subsection{\@startsection{subsection}{2}{\z@}{+2.5ex plus +1ex
minus +.2ex}{1.5ex plus .2ex}{\normalsize\bf\boldmath}}
\def\subsubsection{\@startsection{subsubsection}{3}{\z@}{+3.25ex plus
 +1ex minus +.2ex}{1.5ex plus .2ex}{\normalsize\it}}

\expandafter\ifx\csname mathrm\endcsname\relax\def\mathrm#1{{\rm #1}}\fi

\newcounter{saveeqn}

\@addtoreset{equation}{section}

\newcount\@tempcntc
\def\@citex[#1]#2{\if@filesw\immediate\write\@auxout{\string\citation{#2}}\fi
  \@tempcnta\z@\@tempcntb\m@ne\def\@citea{}\@cite{\@for\@citeb:=#2\do
    {\@ifundefined
       {b@\@citeb}{\@citeo\@tempcntb\m@ne\@citea
        \def\@citea{,\penalty\@m\ }{\bf ?}\@warning
       {Citation `\@citeb' on page \thepage \space undefined}}%
    {\setbox\z@\hbox{\global\@tempcntc0\csname
b@\@citeb\endcsname\relax}%
     \ifnum\@tempcntc=\z@ \@citeo\@tempcntb\m@ne
       \@citea\def\@citea{,\penalty\@m}
       \hbox{\csname b@\@citeb\endcsname}%
     \else
      \advance\@tempcntb\@ne
      \ifnum\@tempcntb=\@tempcntc
      \else\advance\@tempcntb\m@ne\@citeo
      \@tempcnta\@tempcntc\@tempcntb\@tempcntc\fi\fi}}\@citeo}{#1}}

\def\@citeo{\ifnum\@tempcnta>\@tempcntb\else\@citea
  \def\@citea{,\penalty\@m}%
  \ifnum\@tempcnta=\@tempcntb\the\@tempcnta\else
   {\advance\@tempcnta\@ne\ifnum\@tempcnta=\@tempcntb \else
\def\@citea{--}\fi
    \advance\@tempcnta\m@ne\the\@tempcnta\@citea\the\@tempcntb}\fi\fi}

\newcommand{\lsim}
{\mathrel{\raisebox{-.3em}{$\stackrel{\displaystyle <}{\sim}$}}}

\def\asymp#1%
{\mathrel{\raisebox{-.4em}{$\widetilde{\scriptstyle #1}$}}}

\def\Nequal#1%
{\mathrel{\raisebox{-.5em}{$\stackrel{=}{\scriptstyle\rm#1}$}}}
\newcommand{\dsl}[1]{\not \hspace{-0.7mm}#1}
\def\dsl{\mathpalette\make@slash}
\def\make@slash#1#2{\setbox\z@\hbox{$#1#2$}%
  \hbox to 0pt{\hss$#1/$\hss\kern-\wd0}\box0}

\def\beq{\begin{equation}}
\def\eeq{\end{equation}}
\def\beqar{\begin{eqnarray}}
\def\eeqar{\end{eqnarray}}
\def\barr#1{\begin{array}{#1}}
\def\earr{\end{array}}
\def\bfi{\begin{figure}}
\def\efi{\end{figure}}
\def\btab{\begin{table}}
\def\etab{\end{table}}
\def\bce{\begin{center}}
\def\ece{\end{center}}

\def\text{\textstyle}

\def\al{\alpha}

\def\ga{\gamma}
\def\de{\delta}

\def\la{\lambda}

\def\si{\sigma}

\def\reffi#1{\mbox{Figure~\ref{#1}}}

\def\refta#1{\mbox{Table~\ref{#1}}}

\def\refse#1{\mbox{Section~\ref{#1}}}

\def\citere#1{\mbox{Ref.~\cite{#1}}}
\def\citeres#1{\mbox{Refs.~\cite{#1}}}

\newcommand{\TeV}{\unskip\,\mathrm{TeV}}
\newcommand{\GeV}{\unskip\,\mathrm{GeV}}
\newcommand{\MeV}{\unskip\,\mathrm{MeV}}

\newcommand{\ri}{{\mathrm{i}}}

\newcommand{\Oa}{\mathswitch{{\cal{O}}(\alpha)}}

\def\mathswitchr#1{\relax\ifmmode{\mathrm{#1}}\else$\mathrm{#1}$\fi}
\newcommand{\Pf}{\mathswitch  f}

\newcommand{\PW}{\mathswitchr W}
\newcommand{\Pw}{\mathswitchr w}
\newcommand{\PZ}{\mathswitchr Z}

\newcommand{\PH}{\mathswitchr H}
\newcommand{\Pe}{\mathswitchr e}

\newcommand{\Pd}{\mathswitchr d}

\newcommand{\Pu}{\mathswitchr u}

\newcommand{\Ps}{\mathswitchr s}

\newcommand{\Pc}{\mathswitchr c}

\newcommand{\Pb}{\mathswitchr b}

\newcommand{\Pt}{\mathswitchr t}

\newcommand{\Pep}{\mathswitchr {e^+}}
\newcommand{\Pem}{\mathswitchr {e^-}}

\def\mathswitch#1{\relax\ifmmode#1\else$#1$\fi}

\newcommand{\MW}{\mathswitch {M_\PW}}

\newcommand{\MZ}{\mathswitch {M_\PZ}}
\newcommand{\MH}{\mathswitch {M_\PH}}
\newcommand{\Me}{\mathswitch {m_\Pe}}

\newcommand{\Md}{\mathswitch {m_\Pd}}
\newcommand{\Mu}{\mathswitch {m_\Pu}}
\newcommand{\Ms}{\mathswitch {m_\Ps}}
\newcommand{\Mc}{\mathswitch {m_\Pc}}
\newcommand{\Mb}{\mathswitch {m_\Pb}}
\newcommand{\Mt}{\mathswitch {m_\Pt}}
\newcommand{\GW}{\Gamma_{\PW}}
\newcommand{\GZ}{\Gamma_{\PZ}}

\newcommand{\sw}{\mathswitch {s_\Pw}}
\newcommand{\cw}{\mathswitch {c_\Pw}}

\newcommand{\GF}{\mathswitch {G_\mu}}

\def\solid{\raise.9mm\hbox{\protect\rule{1.1cm}{.2mm}}}
\def\dash{\raise.9mm\hbox{\protect\rule{2mm}{.2mm}}\hspace*{1mm}}

\def\ie{i.e.\ }

\newcommand{\IBA}{{\mathrm{IBA}}}
\newcommand{\DPA}{{\mathrm{DPA}}}

\newcommand{\born}{{\mathrm{Born}}}

\newcommand{\eefourf}{{\mathswitchr{ee4f}}}

\newcommand{\CMS}{{\mathrm{CMS}}}
\newcommand{\FW}{{\mathrm{FW}}}

\def\draftdate{\relax}
\def\mda{\relax}
\def\mua{\relax}
\def\mla{\relax}
\def\draft{
\def\thtystars{******************************}
\def\sixtystars{\thtystars\thtystars}
\typeout{}
\typeout{\sixtystars**}
\typeout{* Draft mode!
         For final version remove \protect\draft\space in source file *}
\typeout{\sixtystars**}
\typeout{}
\def\draftdate{\today}
\def\mua{\marginpar[\boldmath\hfil$\uparrow$]%
                   {\boldmath$\uparrow$\hfil}%
                    \typeout{marginpar: $\uparrow$}\ignorespaces}
\def\mda{\marginpar[\boldmath\hfil$\downarrow$]%
                   {\boldmath$\downarrow$\hfil}%
                    \typeout{marginpar: $\downarrow$}\ignorespaces}
\def\mla{\marginpar[\boldmath\hfil$\rightarrow$]%
                   {\boldmath$\leftarrow $\hfil}%
                    \typeout{marginpar: $\leftrightarrow$}\ignorespaces}
\def\Mua{\marginpar[\boldmath\hfil$\Uparrow$]%
                   {\boldmath$\Uparrow$\hfil}%
                    \typeout{marginpar: $\uparrow$}\ignorespaces}
\def\Mda{\marginpar[\boldmath\hfil$\Downarrow$]%
                   {\boldmath$\Downarrow$\hfil}%
                    \typeout{marginpar: $\downarrow$}\ignorespaces}
\def\Mla{\marginpar[\boldmath\hfil$\Rightarrow$]%
                   {\boldmath$\Leftarrow $\hfil}%
                    \typeout{marginpar: $\leftrightarrow$}\ignorespaces}
\overfullrule 5pt
\oddsidemargin -15mm
\marginparwidth 29mm
}

\def\stars{\strut\leaders\hbox{*}\hfill\strut}
\def\starline{\hfil\strut\hfil\hbox to \textwidth {\stars}\hfil}

\begin{document}
\thispagestyle{empty}
\def\thefootnote{\fnsymbol{footnote}}
\setcounter{footnote}{1}
\null
\draftdate\hfill MPP-2005-8 \\
\strut\hfill PSI-PR-05-02\\
\strut\hfill hep-ph/0502063
\vfill
\begin{center}
{\Large \bf\boldmath
Complete electroweak ${\cal O}(\alpha)$ corrections to \\[.5em]
charged-current $\Pep\Pem\to4\,$fermion  processes
\par} \vskip 2.5em
\vspace{1cm}

{\large
{\sc A.\ Denner$^1$, S.\ Dittmaier$^2$, M. Roth$^2$ and 
L.H.\ Wieders$^{1,3}$} } \\[1cm]
$^1$ {\it Paul Scherrer Institut, W\"urenlingen und Villigen\\
CH-5232 Villigen PSI, Switzerland} \\[0.5cm]
$^2$ {\it Max-Planck-Institut f\"ur Physik 
(Werner-Heisenberg-Institut) \\
D-80805 M\"unchen, Germany}
\\[0.5cm]
$^3$ {\it Institute for Theoretical Physics\\ University of Z\"urich, CH-8057 
Z\"urich, Switzerland}
\par \vskip 1em
\end{center}\par
\vskip 1cm {\bf Abstract:} \par 
The complete electroweak ${\cal O}(\alpha)$ corrections are
calculated for the charged-current four-fermion production processes
$\Pep\Pem\to\nu_\tau\tau^+\mu^-\bar\nu_\mu$,
$\Pu\bar\Pd\mu^-\bar\nu_\mu$, and
$\Pu\bar\Pd\Ps\bar\Pc$.
The calculation is performed using complex gauge-boson
masses, supplemented by complex couplings to restore gauge invariance.
The evaluation of the occurring one-loop tensor integrals,
which include 5- and 6-point functions, requires new techniques.
Explicit numerical results are presented for total cross sections
in the energy range from the W-pair-production threshold region up to a 
scattering energy of $2\TeV$. 
A comparison with the predictions based on the 
``double-pole approximation'' (DPA) provided by the generator
{\sc RacoonWW} reveals corrections beyond
DPA of $\lsim0.5\%$ in the energy range $170{-}300\GeV$,
in agreement with previous estimates for the intrinsic DPA uncertainty.
The difference to the DPA increases to $1{-}2\%$ for $\sqrt{s}\sim1{-}2\TeV$.
At threshold, where the DPA becomes unreliable, the full ${\cal O}(\alpha)$
calculation corrects an improved Born approximation (IBA) by about
$1.6\%$, also consistent with an error estimate of the IBA.
\par
\vskip 1cm
\noindent
August 2011     
\null
\setcounter{page}{0}
\clearpage
\def\thefootnote{\arabic{footnote}}
\setcounter{footnote}{0}

\section{Introduction}
\label{se:intro}

At LEP2, W-pair-mediated four-fermion $(4f)$ production was
experimentally explored with quite high precision (see \citere{lep2}
and references therein).  The total W-pair cross section was measured
from threshold up to a centre-of-mass (CM) energy of $207\GeV$;
combining the cross-section measurements a precision of $\sim 1\%$ was
reached. The W-boson mass $\MW$ was determined from the threshold
cross section with an error of $\sim 200\MeV$ and by reconstructing
the $\PW$ bosons from their decay products within $\sim 40\MeV$.
Deviations from the Standard Model (SM) triple gauge-boson couplings,
usually quantified in the parameters $\Delta g^\PZ_1$,
$\Delta\kappa_\gamma$, and $\lambda_\gamma$, were constrained within a
few per cent.

The LEP2 measurements had set the scale in accuracy in the theoretical
\looseness -1
predictions for W-pair-mediated $4f$ production.  The theoretical
treatment and the gained level in precision are reviewed in
\citeres{Beenakker:1996kt,Grunewald:2000ju}.  The $\PW$ bosons are
treated as resonances in the full $4f$ processes, $\Pe^+ \Pe^- \to 4
\Pf\, (+\,\gamma)$.  Radiative corrections are split into universal
and non-universal corrections.  The former comprise
leading-logarithmic corrections from initial-state radiation (ISR),
higher-order corrections included by using appropriate effective
couplings, and the Coulomb singularity.  These corrections can be
combined with the 
lowest-order matrix elements easily.  The
remaining corrections are called non-universal, since they depend on
the process under investigation. For LEP2 accuracy, it was sufficient
to include these corrections in the so-called double-pole
approximation (DPA), where only the leading term in an expansion about
the poles in the two W-boson propagators is taken into account.
Different versions of such a DPA have been used in the literature
\cite{Beenakker:1998gr,Jadach:1998tz,Jadach:2000kw,Denner:2000kn,Kurihara:2001um}.
Although several Monte Carlo programs exist that include universal
corrections, only two event generators, {\sc YFSWW}
\cite{Jadach:1998tz,Jadach:2000kw} and {\sc RacoonWW}
\cite{Denner:2000kn,Denner:1999gp,Denner:2001zp}, include
non-universal corrections.

In the DPA approach, the W-pair cross section can be predicted within
$\sim0.5\%$ $(0.7\%)$ in the energy range between $180\GeV$ ($170\GeV$)
and $\sim 500\GeV$, which was sufficient for the LEP2 accuracy of
$\sim 1\%$ for energies $170{-}207\GeV$. In the threshold region
($\sqrt{s}\lsim 170\GeV$), the DPA is not reliable, and the best
available prediction results from an improved Born approximation (IBA)
based on leading universal corrections only, and thus possesses an
intrinsic uncertainty of $\sim 2\%$.  At energies above $500\GeV$
effects beyond ${\cal O}(\alpha)$, such as Sudakov logarithms at
higher orders, become important and should be included in predictions
at per-cent accuracy.

At a future International $\Pe^+ \Pe^-$ Linear Collider (ILC)
\cite{Aguilar-Saavedra:2001rg,Abe:2001wn,Abe:2001gc}, the accuracy of
the cross-section measurement will be at the per-mille level, and the
precision of the $\PW$-mass determination is expected to be $\sim
10\MeV$ \cite{talkKM} by direct reconstruction and \mbox{$\sim 7\MeV$}
from a threshold scan of the total W-pair-production cross section
\cite{Aguilar-Saavedra:2001rg,Abe:2001wn}.  The theoretical
uncertainty (TU) for the direct mass reconstruction at LEP2 is
estimated to be of the order of $\sim 5\MeV$ \cite{Jadach:2001cz} to
$\lsim 10\MeV$ \cite{Cossutti}, based on results of
{\sc YFSWW} and {\sc
  RacoonWW}; theoretical improvements are, thus, desirable for an ILC.
For the cross-section prediction at threshold the TU is only $\sim 2\%$,
because it is based on an IBA, and 
thus is definitely insufficient for the planned
precision measurement of $\MW$ in a threshold scan.  The main
sensitivity of all observables to anomalous couplings in the
triple gauge-boson vertices is provided by the W-pair production angle
distribution.  The TU in constraining the parameter $\la_\gamma$ was
estimated to be $\sim 0.005$ \cite{Bruneliere:2002df} for the LEP2
analysis. Since a future ILC is more sensitive to anomalous 
gauge-boson couplings
than LEP2 by more than an order of magnitude, a further reduction of
the uncertainties 
resulting from missing radiative corrections is necessary. In
summary, these considerations demonstrate the necessity of a full
one-loop calculation for the $\Pep\Pem\to 4f$ process and of further
improvements by leading higher-order corrections.

In this paper we present first results of a complete
${\cal O}(\alpha)$ calculation (improved by higher-order ISR)
for the $4f$ final states 
$\nu_\tau\tau^+\mu^-\bar\nu_\mu$,
$\Pu\bar\Pd\mu^-\bar\nu_\mu$, and
$\Pu\bar\Pd\Ps\bar\Pc$,
which are relevant for W-pair production.%
\footnote{Electrons and/or positrons in the final state are not yet considered;
they deserve further refinements, in particular the inclusion of
finite-electron-mass effects in the domain of forward-scattered
$\Pe^\pm$. We also do not yet include final states that can also be
produced via two resonant Z~bosons, so called mixed CC/NC reactions.
These can be taken into account in lowest order in RacoonWW.}
The actual calculation is rather complicated.%
\footnote{Some of the problems appearing in a first attempt of such a
  calculation were already described in \citere{Vicini:1998iy}.
  Recently the authors of the {\sc GRACE/1-LOOP} system reported on
  progress towards a full one-loop calculation for
  $\Pep\Pem\to\mu^-\bar\nu_\mu\Pu\bar\Pd$ in \citere{Boudjema:2004id}
  so that one can expect that the system will be able to deal with
  $\Pep\Pem\to4f$ processes at one loop in the near future.}
Technically the occurring one-loop tensor integrals comprise 5- and
6-point functions up to rank~3, and conceptually the W-boson resonances
require a treatment in loop diagrams that preserves gauge invariance.
Details of our approach will be described in a forthcoming
publication; in the next section we just sketch the most important
features of the calculation.  In \refse{se:numres} we present explicit
numerical results on total cross sections for scattering energies from
near the W-pair-production threshold up to $2\TeV$.  Particular
attention is paid to a comparison with the DPA and IBA approaches used
at LEP2.

\section{Method of calculation}
\label{se:calc}

\begin{sloppypar}
The actual calculation builds upon the {\sc RacoonWW} approach
\cite{Denner:2000kn}, where real-photonic corrections are based on
full matrix elements and virtual corrections are treated in DPA. Real
and virtual corrections are combined either using two-cutoff
phase-space slicing or employing the dipole subtraction method
\cite{Dittmaier:2000mb,Roth:1999kk} for photon radiation.  
We also include leading-logarithmic initial-state radiation (ISR)
beyond $\Oa$ in the structure-function approach
(\citere{Beenakker:1996kt} and references therein). 
The presented calculation only differs in the treatment of the
virtual corrections from {\sc RacoonWW}. 
We neglect the masses of the fermions whenever possible, \ie
everywhere but in the mass-singular logarithms,
and set the quark-mixing matrix to the unit matrix.

In the following we sketch the main difficulties in the calculation
and briefly explain our solutions.
\end{sloppypar}

\subsection{Technical issues}

In contrast to the DPA approach, the one-loop calculation of
an $\Pep\Pem\to4f$ process requires the evaluation of 5- and 6-point
one-loop tensor integrals. Some 6-point diagrams are shown in
\reffi{fig:hexadiags} for illustration. 
\bfi
{\small\unitlength0.9pt \SetScale{.9}
\centerline{
\begin{picture}(490,100)(0,0)
\put(0,0){
\begin{picture}(150,100)(0,-50)
\ArrowLine(40,20)(10,20)
\ArrowLine(40,-20)(40,20)
\ArrowLine(10,-20)(40,-20)
\ArrowLine(130,15)(100,15)\ArrowLine(100,15)(80,40)
\ArrowLine(80,40)(130,40)
\ArrowLine(100,-15)(130,-15)\ArrowLine(80,-40)(100,-15)
\ArrowLine(130,-40)(80,-40)
\Photon(100,-15)(100,15){2}{3.5}
\Photon(40,20)(80,40){2}{4.5}
\Photon(40,-20)(80,-40){-2}{4.5}
\Vertex(100,-15){2}\Vertex(100,15){2}
\Vertex(40,20){2}\Vertex(80,40){2}
\Vertex(40,-20){2}\Vertex(80,-40){2}
\Text(6,20)[r]{$e^+$}
\Text(6,-20)[r]{$e^-$}
\Text(35,0)[r]{$\nu_e$}
\Text(135,40)[l]{$f_1$}
\Text(135,15)[l]{$\bar f_2$}
\Text(135,-15)[l]{$f_3$}
\Text(135,-40)[l]{$\bar f_4$}
\Text(98,27)[l]{$f_2$}
\Text(98,-27)[l]{$f_3$}
\Text(60,40)[b]{$W$}
\Text(60,-40)[t]{$W$}
\Text(105,0)[l]{$\ga/Z$}
\end{picture}}
\put(170,0){
\begin{picture}(150,100)(0,-50)
\ArrowLine(40,20)(10,20)
\ArrowLine(40,-20)(40,20)
\ArrowLine(10,-20)(40,-20)
\ArrowLine(130,15)(100,15)\ArrowLine(100,15)(80,40)
\ArrowLine(80,40)(130,40)
\ArrowLine(100,-15)(130,-15)\ArrowLine(80,-40)(100,-15)
\ArrowLine(130,-40)(80,-40)
\Photon(100,-15)(100,15){2}{3.5}
\Photon(40,20)(80,40){2}{4.5}
\Photon(40,-20)(80,-40){-2}{4.5}
\Vertex(100,-15){2}\Vertex(100,15){2}
\Vertex(40,20){2}\Vertex(80,40){2}
\Vertex(40,-20){2}\Vertex(80,-40){2}
\Text(6,20)[r]{$e^+$}
\Text(6,-20)[r]{$e^-$}
\Text(35,0)[r]{$e$}
\Text(135,40)[l]{$f_1$}
\Text(135,15)[l]{$\bar f_2$}
\Text(135,-15)[l]{$f_3$}
\Text(135,-40)[l]{$\bar f_4$}
\Text(60,40)[b]{$\ga/Z$}
\Text(60,-40)[t]{$\ga/Z$}
\Text(105,0)[l]{$W$}
\Text(98,27)[l]{$f_1$}
\Text(98,-27)[l]{$f_4$}
\end{picture}}
\put(340,0){
\begin{picture}(150,100)(0,-50)
\ArrowLine(40,-20)(10,-20)
\ArrowLine(40,20)(40,-20)
\ArrowLine(10,20)(40,20)
\ArrowLine(130,15)(100,15)\ArrowLine(100,15)(80,40)
\ArrowLine(80,40)(130,40)
\ArrowLine(100,-15)(130,-15)\ArrowLine(80,-40)(100,-15)
\ArrowLine(130,-40)(80,-40)
\Photon(100,-15)(100,15){2}{3.5}
\Photon(40,20)(80,40){2}{4.5}
\Photon(40,-20)(80,-40){-2}{4.5}
\Vertex(100,-15){2}\Vertex(100,15){2}
\Vertex(40,20){2}\Vertex(80,40){2}
\Vertex(40,-20){2}\Vertex(80,-40){2}
\Text(6,-20)[r]{$e^+$}
\Text(6,20)[r]{$e^-$}
\Text(35,0)[r]{$e$}
\Text(135,40)[l]{$f_1$}
\Text(135,15)[l]{$\bar f_2$}
\Text(135,-15)[l]{$f_3$}
\Text(135,-40)[l]{$\bar f_4$}
\Text(60,40)[b]{$\ga/Z$}
\Text(60,-40)[t]{$\ga/Z$}
\Text(105,0)[l]{$W$}
\Text(98,27)[l]{$f_1$}
\Text(98,-27)[l]{$f_4$}
\end{picture}}
\end{picture}}
}
\caption{Ten 6-point diagrams contributing to $\Pep\Pem\to f_1 \bar f_2 f_3 \bar f_4$. The remaining 30  6-point diagrams are obtained
  by reversing the fermion flow in one or both of the fermion chains
  corresponding to the outgoing fermions.}
\label{fig:hexadiags}
\efi
For the generic $f_1 \bar f_2 f_3 \bar f_4$ final state, where $f_1$
and $f_3$ are different fermions excluding electrons and electron
neutrinos and $f_2$ and $f_4$ their isospin partners, there are 40
hexagon diagrams, 112 pentagon diagrams, and 227 (220) box diagrams in
the conventional 't Hooft--Feynman gauge (background-field gauge).
We calculate the 6-point integrals by directly reducing them to six
5-point functions, as described in \citeres{Me65,Denner:1993kt}.  The
5-point integrals are reduced to five 4-point functions following the
method of \citere{Denner:2002ii}.  These reduction steps involve
so-called (modified) Cayley determinants, the zeroes of which are
related to the Landau singularities of the (sub-)diagrams. We did not
encounter numerical problems with these determinants.  Note that this
reduction of 5- and 6-point integrals to 4-point integrals does not
involve inverse Gram determinants composed of external momenta, which
naturally occur in the Passarino--Veltman reduction
\cite{Passarino:1979jh} of tensor to scalar integrals. The latter
procedure leads to serious numerical problems when the Gram
determinants become small, which happens usually near the boundary of
phase space but can also occur within phase space because of the
indefinite Minkowski metric.

We use, however, Passarino--Veltman reduction to calculate tensor
\looseness -1
integrals up to 4-point functions, which involves inverse Gram
determinants built from two or three momenta.  This, in fact, leads to
numerical instabilities in phase-space regions where these Gram
determinants become small.  For these regions we have worked out two
``rescue systems'': one makes use of expansions of the tensor
coefficients about the limit of vanishing Gram determinants; in the
other, alternative method we numerically evaluate a specific tensor
coefficient, the integrand of which is logarithmic
in Feynman parametrization, and derive the remaining
coefficients as well as the scalar integral from it algebraically.
This reduction again involves only inverse Cayley determinants, but
no inverse Gram determinants.

In addition to the evaluation of the one-loop integrals, also the
evaluation of the three spinor chains corresponding to the
three external fermion--antifermion pairs is non-trivial, because
the chains are contracted with each other and/or with 
four-momenta in many different ways. There are ${\cal O}(10^3)$
different chains to calculate, so that an algebraic reduction to a
standard form which involves only very few standard chains is
desirable.  We have worked out algorithms that reduce all
occurring spinor chains to ${\cal O}(10)$ standard structures without
introducing coefficients that lead to numerical problems.

\subsection{Conceptual issues}

The description of resonances in (standard) perturbation theory 
requires a Dyson summation of self-energy insertions in the resonant
propagator in order to introduce the imaginary part provided by the
finite decay width into the propagator denominator.
It is well known that this procedure in general violates gauge
invariance, \ie destroys Slavnov--Taylor or Ward identities
and disturbs the cancellation of gauge-parameter dependences,
because different perturbative orders are mixed
(see, for instance, \citere{Grunewald:2000ju} and references therein).
Several solutions have been described for lowest-order predictions,
but the general problem is still considered as unsolved for a 
consistent evaluation of radiative corrections.
The DPA provides a gauge-invariant answer in terms of an expansion about the
resonance,%
\footnote{The recently proposed approach to describe unstable
  particles within an effective field theory \cite{Beneke:2003xh} is
  equivalent to a pole expansion.}  but in the full calculation we are
after a unified description that is valid both for resonant and
non-resonant regions in phase space, without any matching between
different treatments for different regions.

In our calculation we use a generalization of the so-called
``complex-mass scheme'' (CMS), which was introduced in
\citere{Denner:1999gp} for lowest-order calculations, to the one-loop
level.  In this approach the W- and Z-boson masses are consistently
considered as complex quantities, defined as the locations of the
poles in the complex $p^2$ plane of the corresponding propagators with
momentum $p$.  Gauge invariance is preserved if the complex masses are
introduced everywhere in the Feynman rules, in particular, in the
definition of the weak mixing angle,
\beq
\cw^2 = 1-\sw^2 = \frac{\MW^2-\ri\MW\GW}{\MZ^2-\ri\MZ\GZ},
\eeq
which is now derived from the ratio of the complex mass squares.
The (algebraic) relations, such as Ward identities, that follow from gauge
invariance remain valid, because the gauge-boson masses are modified
only by an analytic continuation. Since this continuation already
modifies the lowest-order predictions by changing the gauge-boson
masses, double-counting of higher-order effects has to be carefully
avoided by an appropriate renormalization procedure.

\begin{sloppypar}
The use of complex gauge-boson masses necessitates the consistent use
of these complex masses also in loop integrals.  To this end, we have
derived all relevant one-loop integrals with complex internal masses.
The IR-singular integrals were taken from \citere{Beenakker:1990jr}.
Concerning the non-IR singular cases, we have analytically continued
the results of \citere{'tHooft:1979xw} for the 2-point and 3-point
functions\footnote{Note that the result of \citere{'tHooft:1979xw} for
  the scalar two-point function is not valid in general for complex
  masses. In this case an extra $\eta$ function has to be added. The
  same comment applies to the results for the 2-point tensor integrals
  in
  \citere{Passarino:1979jh}.}%
, and the relevant results of
\citere{Denner:1991qq} for the 4-point functions. We have checked all
these results by independent direct calculation of the
Feynman-parameter integrals.
\end{sloppypar}

\subsection{Checks}

The complexity of the calculation enforces a number of consistency checks
to prove the reliability of the results. We have performed the 
following checks:
\begin{itemize}
\item
{\it UV finiteness} is checked by verifying the independence of the
(arbitrary) reference scale $\mu$ of dimensional regularization.
\item {\it IR finiteness} is checked by varying the logarithm
  $\ln\lambda$ of the (formally infinitesimal) photon mass $\la$,
  which leaves the sum of the virtual and the soft-photonic
  corrections (slicing approach) or of the virtual and endpoint
  contributions (\ie the singular parts that are subtracted from the
  virtual corrections in the subtraction approach) invariant.
\item
{\it Mass singularities} related to collinear photon emission or
exchange are checked by verifying the independence of the sum
of the virtual corrections and the subtraction endpoint contributions
from the small masses of the external fermions.
\item
{\it Gauge invariance} is checked by comparing the result obtained
within the 't~Hooft--Feynman gauge with an independent result
obtained within the background-field formalism \cite{Denner:1994xt}.
Apart from diagrams that involve only fermion--gauge-boson couplings
in the loops, the contributions of individual Feynman graphs
are in general different in the two approaches.
\item
{\it The real corrections} are taken over from {\sc RacoonWW}
\cite{Denner:2000kn,Denner:1999gp}, where they were checked by two
independent calculations in detail. 
Moreover, agreement was found between the results obtained with
phase-space slicing or dipole subtraction.
\item {\it The scalar loop integrals} for complex masses have been
  calculated in two completely independent ways and implemented in two
  independent in-house libraries. We checked that these results agree
  with each other and in the limit of zero width also with FF
  \cite{vanOldenborgh:1990yc}.
\item {\it Two completely independent} calculations have been
  performed within our group, revealing good agreement. All algebraic
  manipulations, including the generation of Feynman diagrams and the
  reduction of amplitudes to standard forms, have been done using two
  independent programs.%
  \footnote{The amplitudes are generated with {\sc FeynArts}, using
  the two independent versions 1 and 3, as described in 
  \citeres{Kublbeck:1990xc} and \cite{Hahn:2000kx}, respectively.
  The algebraic manipulations are performed using two independent
  in-house programs implemented in {\sc Mathematica}, one of which
\looseness-1
  builds upon {\sc FormCalc}~\cite{Hahn:1998yk}.}
  The evaluations of all scalar and tensor loop
  integrals are based on two independent in-house libraries, which
  employ the two different rescue systems mentioned above.  We
  consider this as the most important and convincing check.
\end{itemize}
We emphasize that all these checks, including the gauge-invariance check,
have been carried out for non-zero gauge-boson widths.

\section{Numerical results}
\label{se:numres}

\subsection{Input parameters and setup}
\label{se:setup}

For the numerical evaluation we use the following set of
SM parameters,
\beq\arraycolsep 2pt
\begin{array}[b]{lcllcllcl}
\GF & = & 1.16637 \times 10^{-5} \GeV^{-2}, \quad&
\alpha(0) &=& 1/137.03599911, &
\alpha_{\mathrm{s}} &=& 0.1187,\\
\MW & = & 80.425\GeV, &
\MZ & = & 91.1876\GeV, &
\GZ & = & 2.4952\GeV, \\
\MH & = & 115\GeV, \\
\Me & = & 0.51099892\MeV, &
m_\mu &=& 105.658369\MeV,\quad &
m_\tau &=& 1.77699\GeV, \\
\Mu & = & 66\MeV, &
\Mc & = & 1.2\GeV, &
\Mt & = & 178\;\GeV, 
\\
\Md & = & 66\MeV, &
\Ms & = & 150\MeV, &
\Mb & = & 4.3\GeV, 
\end{array}
\label{eq:SMpar}
\eeq
which essentially follows \citere{Eidelman:2004wy}.  For the top-quark
mass $\Mt$ we have taken the more recent value of
\citere{Azzi:2004rc}.  The masses of the light quarks are adjusted to
reproduce the hadronic contribution to the photonic vacuum
polarization of \citere{Jegerlehner:2001ca}. Since we parametrize the
lowest-order cross section with the Fermi constant $\GF$ ($\GF$
scheme), \ie we derive the electromagnetic coupling $\alpha$ according
to $ \alpha_{\GF} = \sqrt{2}\GF\MW^2(1-\MW^2/\MZ^2)/\pi$, the results are
practically independent of the masses of the light quarks.  Moreover,
this procedure absorbs the corrections proportional to $\Mt^2/\MW^2$
in the fermion--W-boson couplings and the running of $\al(Q^2)$ from
$Q^2=0$ to the electroweak scale. In the relative radiative
corrections, we use, however, $\alpha(0)$ as coupling parameter, which
is the correct effective coupling for real photon emission.

QCD corrections are treated in the ``naive'' approach of multiplying cross
sections and partial decay rates by factors
$(1+\alpha_{\mathrm{s}}/\pi)$ per hadronically decaying W~boson.  The
W-boson width $\GW$ is calculated from the above input including
electroweak ${\cal O}(\alpha)$ and QCD corrections, yielding
\beq
\GW = 2.09269848\ldots\GeV.
\eeq
For the full one-loop calculation and for the DPA approach this
procedure ensures that the effective branching ratios for the leptonic,
semileptonic, and hadronic W~decays, which result from the
integration over the decay fermions, add up to 1.
In order to keep this normalization also for the IBA, $\GW$ is
calculated in the corresponding approximation, \ie in the
$\GF$ scheme without electroweak corrections, yielding
\beq
\GW\Big|_{\IBA} = 2.10009936\ldots\GeV.
\eeq
A detailed description of the IBA, as used in {\sc RacoonWW}, can
be found in \citere{Denner:2001zp}.

In the following we discuss only total cross sections without
any phase-space cuts.
The presented results have been obtained
with $10^7$ events, using the subtraction method.

\subsection{Results for total cross sections}

Tables~\ref{tab:cstot_lept}--\ref{tab:cstot_had} show some
representative results on total cross sections for the final
states $\nu_\tau\tau^+\mu^-\bar\nu_\mu$,
$\Pu\bar\Pd\mu^-\bar\nu_\mu$, and $\Pu\bar\Pd\Ps\bar\Pc$
in various approximations for different CM energies $\sqrt{s}$.
The numbers in parentheses represent the uncertainties in the last
digits of the predictions.
\begin{table}[tb]
$\begin{array}{rccccc}
\rlap{$\Pep\Pem\to\nu_\tau\tau^+\mu^-\bar\nu_\mu$}
\phantom{\sqrt{s}/\mathrm{GeV}}
\\
\hline
\sqrt{s}/\mathrm{GeV} & \born(\FW) & \born(\CMS) & 
\IBA & \DPA & \eefourf 
\\
\hline
161  &  50.04(2) &  50.01(2) &  37.18(2) &  37.08(2) & 38.16(3)
\\[-.3em]
     &              & [-0.06\%] &[-25.67(6)\%]&[-25.90(3)\%]& [-23.75(4)\%] 
\\
170  & 160.53(6)  & 160.44(6)  & 129.12(6) & 129.17(6) & 130.01(6)
\\[-.3em]
     &              & [-0.06\%] &[-19.52(5)\%]&[-19.53(3)\%]& [-19.01(3)\%]
\\
189  & 216.57(8)  & 216.45(8)  & 191.89(8) & 191.66(9) & 191.92(9)
\\[-.3em]
     &              & [-0.06\%] &[-11.35(5)\%]&[-11.50(2)\%]&[-11.33(3)\%]
\\
200  & 220.41(9)  & 220.29(9)  & 201.13(9) & 200.04(10) & 200.25(10)
\\[-.3em]
     &              & [-0.06\%] & [-8.70(6)\%]& [-9.24(2)\%]& [-9.15(3)\%]
\\
500  &  86.95(5)  & 86.90(5)   &  92.79(5) &  89.81(6) &  89.53(6)
\\[-.3em]
     &              & [-0.06\%] & [+6.78(9)\%]& [+3.29(3)\%]& [+2.97(4)\%]
\\
1000 &  33.35(2) & 33.33(2)  &  38.04(4) &  35.76(3) & 35.53(3)
\\[-.3em]
     &             & [-0.06\%]  &[+14.12(14)\%]& [+7.21(5)\%]& [+6.51(6)\%]
\end{array}$
\caption{Total cross sections in fb for 
$\Pep\Pem\to\nu_\tau\tau^+\mu^-\bar\nu_\mu$
in Born approximation (in the fixed-width and complex-mass schemes),
IBA, DPA, and using the full ${\cal O}(\alpha)$
correction (\eefourf); all but the Born cross sections include
higher-order ISR corrections.}
\label{tab:cstot_lept}
\end{table}%

\begin{figure}
{\unitlength 1cm
\begin{picture}(16,15.5)
\put(-4.6,- 7.2){\includegraphics{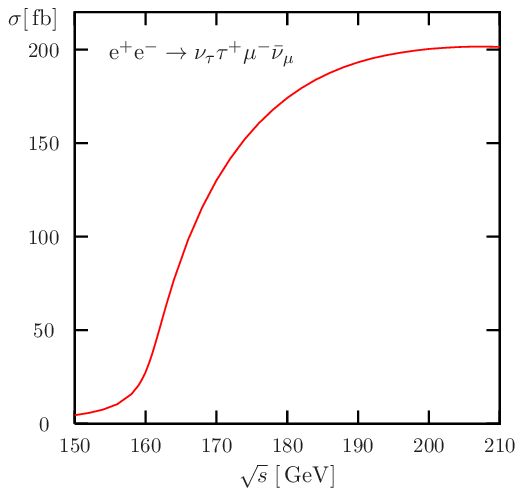}}
\put( 3.6,- 7.2){\includegraphics{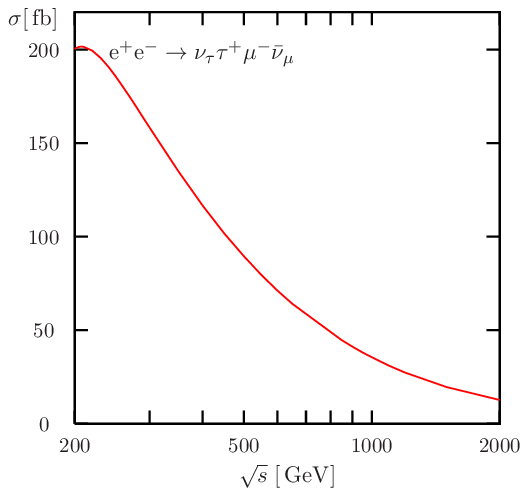}}
\put(-4.6,-15.2){\includegraphics{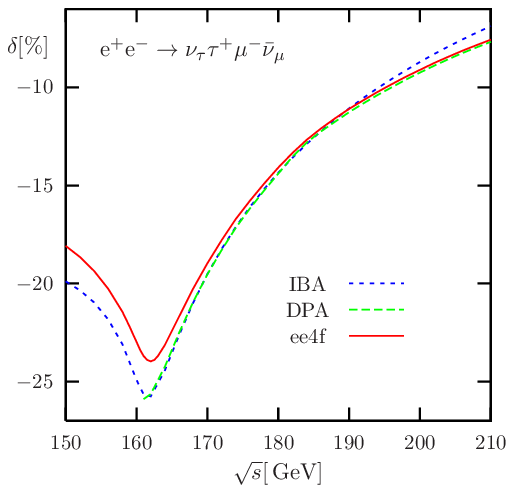}}
\put( 3.6,-15.2){\includegraphics{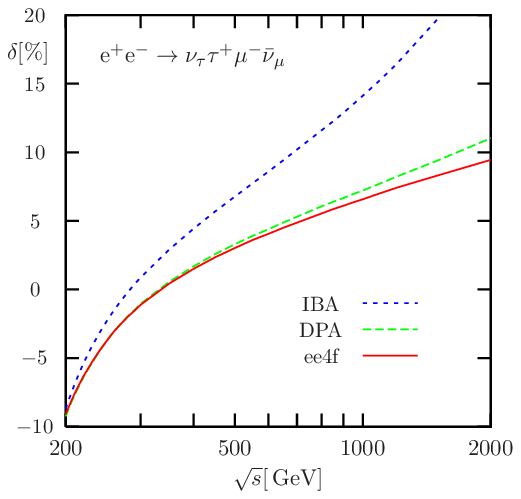}}
\end{picture} }
\caption{Absolute cross section $\si$
(upper plots) and relative corrections $\de$
(lower plots), as defined in the text,
to the total cross section without cuts for
$\Pep\Pem\to\nu_\tau\tau^+\mu^-\bar\nu_\mu$ obtained from the
IBA, DPA, and the full ${\cal O}(\alpha)$ calculation (\eefourf). All
predictions are improved by higher-order ISR.}
\label{fig:leptrelrc}
\end{figure}

\begin{table}
$\begin{array}{rccccc}
\rlap{$\Pep\Pem\to\Pu\bar\Pd\mu^-\bar\nu_\mu$}
\phantom{\sqrt{s}/\mathrm{GeV}}
\\
\hline
\sqrt{s}/\mathrm{GeV} & \born(\FW) & \born(\CMS) & 
\IBA & \DPA & \eefourf 
\\
\hline
161  & 150.15(7) & 150.07(7) & 115.75(7) & 115.48(7) & 118.77(8)
\\[-.3em]
     &             &[-0.06\%] &[-22.87(6)\%]&[-23.09(3)\%]&[-20.86(4)\%]
\\
170  & 481.6(2) & 481.3(2)  & 402.0(2)  & 401.8(2)  & 404.5(2)
\\[-.3em]
     &             &[-0.06\%] &[-16.48(5)\%]&[-16.58(3)\%]& [-15.96(3)\%]
\\
189  & 649.7(3) & 649.4(3)  & 597.4(3)  & 596.1(3)  & 597.0(3)
\\[-.3em]
     &             &[-0.06\%] &[-8.00(5)\%] & [-8.26(3)\%]&[-8.06(3)\%]
\\
200  & 661.3(3) & 660.9(3)  & 626.2(3)  & 622.2(3)  & 622.9(3)
\\[-.3em]
     &             &[-0.06\%] &[-5.26(6)\%] &[-5.91(3)\%] &[-5.75(3)\%]
\\
500  & 260.9(1) & 260.8(1)  & 288.9(2)  & 279.6(2)  & 278.6(2)
\\[-.3em]
     &             &[-0.06\%] &[+10.78(9)\%]&[+7.14(4)\%] &[+6.84(4)\%]    
\\
1000 & 100.10(6) &  100.04(6) & 118.44(13) & 111.45(9) & 110.65(10)
\\[-.3em]
     &             &[-0.06\%] &[+18.39(15)\%]&[+11.34(5)\%] &[+10.60(7)\%]
\end{array}$
\caption{Total cross sections in fb for 
$\Pep\Pem\to\Pu\bar\Pd\mu^-\bar\nu_\mu$
in Born approximation (in the fixed-width and complex-mass schemes),
IBA, DPA, and using the full ${\cal O}(\alpha)$
correction (\eefourf); all but the Born cross sections include
higher-order ISR and QCD corrections.}
\label{tab:cstot_semilept}
\end{table}%
\begin{table}
$\begin{array}{rccccc}
\rlap{$\Pep\Pem\to\Pu\bar\Pd\Ps\bar\Pc$}
\phantom{\sqrt{s}/\mathrm{GeV}}
\\
\hline
\sqrt{s}/\mathrm{GeV} & \born(\FW) & \born(\CMS) & 
\IBA & \DPA & \eefourf 
\\
\hline
161  &  450.5(2) &  450.3(2) &  360.4(2) &  359.7(2) &  369.5(3)
\\[-.3em]
     &              &[-0.06\%]  &[-19.97(6)\%]&[-20.16(4)\%]&[-17.94(4)\%]
\\
170  & 1444.9(5)  & 1444.1(5)  & 1251.6(5) & 1250.1(6) & 1258.3(6)
\\[-.3em]
     &              &[-0.06\%]  &[-13.33(5)\%]&[-13.49(3)\%]&[-12.87(3)\%]
\\
189  & 1949.3(8)  & 1948.2(8)  & 1860.0(8) & 1853.8(9) & 1856.9(9)
\\[-.3em]
     &              &[-0.06\%]  &[-4.53(6)\%] &[-4.90(3)\%] &[-4.69(3)\%]
\\
200  & 1983.9(8)  & 1982.9(8)  & 1949.5(9) & 1935.3(9) & 1937.8(10)
\\[-.3em]
     &              &[-0.06\%]  & [-1.68(6)\%]&[-2.45(3)\%] &[-2.27(3)\%]
\\
500  &  782.9(4)  &  782.5(4)  &  899.4(5) &  869.4(6) & 866.7(6)
\\[-.3em]
     &              &[-0.06\%]  &[+14.94(9)\%]&[+11.05(4)\%]&[+10.76(4)\%]
\\
1000 &  300.3(2) & 300.1(2)  &  368.7(4) &  346.1(3) & 343.6(3)
\\[-.3em]
     &             &[-0.06\%]  &[+22.86(16)\%]&[+15.26(5)\%]&[+14.49(7)\%]
\end{array}$
\caption{As in \refta{tab:cstot_semilept}, but for the process 
$\Pep\Pem\to\Pu\bar\Pd\Ps\bar\Pc$.}
\label{tab:cstot_had}
\end{table}%
Columns two and three in each table contain the two versions of the
lowest-order cross section for the full $\Pep\Pem\to4f$ processes
corresponding to the different treatments
of finite-width effects as provided by the ``fixed-width scheme'' (FW)
and the complex-mass scheme (CMS). In the FW scheme the finite
constant width, and thus the complex mass, is only inserted into the
resonant propagators. 
The relative difference $\si_{\born}(\CMS)/\si_{\born}(\FW)-1$ of the
schemes in lowest order is given by the numbers in square brackets in
the third columns. 
Note that the difference is only $0.06\%$ for the
considered energies, so that it is not essential to which Born cross
section we normalize relative corrections.  We have not given an error
on this difference, because the two Born predictions are strongly
correlated. The fourth columns show the IBA \cite{Denner:2001zp}
implemented in {\sc RacoonWW}, which is
based on universal corrections only and includes solely the
contributions of the CC03 diagrams; the numbers in square brackets are
defined as $\de_{\IBA} = \si_{\IBA}/\si_{\born}(\CMS)-1$.
The fifth columns show the DPA of {\sc RacoonWW}%
\footnote{We recall that the DPA of {\sc RacoonWW} goes beyond a pure
pole approximation in two respects. The real-photonic corrections
are based on the full $\Pep\Pem\to4f+\gamma$ matrix elements, and the
Coulomb singularity is included for off-shell W~bosons. 
Further details can be found in \citere{Denner:2000kn}.},
which comprises also
non-universal corrections \cite{Denner:2000kn};
the numbers in square brackets are defined as
$\de_{\DPA} = \si_{\DPA}/\si_{\born}(\FW)-1$.
We normalize $\si_{\DPA}$ to $\si_{\born}(\FW)$, because the lowest-order
part of the DPA is per default evaluated in the FW scheme in {\sc RacoonWW}.
Finally, the last columns (\eefourf) contain the full one-loop corrections
to $\Pep\Pem\to4f$; the numbers in square brackets are defined as
$\de_{\eefourf} = \si_{\eefourf}/\si_{\born}(\CMS)-1$.
Here we normalize to $\si_{\born}(\CMS)$, because the full $\Pep\Pem\to4f$
calculation is consistently performed in the CMS.
Note that all but the ``Born'' numbers include improvements by ISR
effects beyond ${\cal O}(\alpha)$, as described in \citere{Denner:2000kn}.
Additionally the results on the semileptonic and hadronic cross sections
(all but the Born cross results) include naive 
QCD corrections, as explained in \refse{se:setup}.
For better illustration \reffi{fig:leptrelrc} depicts the predictions
for the energy ranges of LEP2 and of the high-energy phase of a future ILC,
focusing on the leptonic final state $\nu_\tau\tau^+\mu^-\bar\nu_\mu$.
The respective figures for the relative corrections $\de$ to the
semileptonic and hadronic final states look almost identical, up to an
offset resulting from the QCD corrections.

A comparison between the DPA and the predictions based on the full
${\cal O}(\alpha)$ corrections reveals differences in the relative
corrections $\de$ of $\lsim 0.3\%\; (0.6\%)$ for CM energies ranging
from $\sqrt{s}\sim200\GeV$ ($170\GeV$) to $500\GeV$.  This is in
agreement with the expected reliability of the DPA, as discussed in
\citeres{Grunewald:2000ju,Jadach:2000kw,Denner:2000kn}.  At higher
energies, the deviations increase and reach $0.7{-}1.6\%$ at
$\sqrt{s}=1{-}2\TeV$.  In the threshold region
($\sqrt{s}\lsim170\GeV$), as expected, the DPA also becomes worse
w.r.t.\ the full one-loop calculation, because the naive error
estimate of $({\alpha}/{\pi})\times({\GW}/{\MW})$ times some numerical
safety factor of ${\cal O}(1{-}10)$ for the corrections missing in the
DPA has to be replaced by
$({\alpha}/{\pi})\times{\GW}/{(\sqrt{s}-2\MW})$ in the threshold
region and thus becomes large.  In view of that, the DPA is even
surprisingly good near threshold.  For CM energies below $170\GeV$ the
LEP2 cross section analysis was based on approximations like the shown
IBA, which follows the full one-loop corrections even below the
threshold at $\sqrt{s}=2\MW$ within an accuracy of about $2\%$, as
expected in \citere{Denner:2001zp}.  The shape of the relative
corrections in the threshold region is determined by ISR. The minimum
in the relative corrections is correlated to the maximum in the slope
of the total cross section.

A detailed comparison \cite{Grunewald:2000ju}
between the DPA versions of \citere{Beenakker:1998gr}, of {\sc YFSWW}
\cite{Jadach:1998tz,Jadach:2000kw}, and of {\sc RacoonWW}
\cite{Denner:2000kn,Denner:1999gp,Denner:2001zp} showed differences of
the same order 
as the differences between the DPA of {\sc
  RacoonWW} and the result based on the full one-loop calculation of
$\Pep\Pem\to4f$ discussed above.  Therefore, 
these DPA predictions are
consistent with the results based on the full one-loop calculation.
The DPA version presented in \citere{Kurihara:2001um}, however,
deviates from the full one-loop calculation by about $0.5\%$ for
typical LEP2 energies.

In addition to the above results which include ISR beyond ${\cal O}(\alpha)$
and QCD corrections,
we present also explicit numbers that include only the
genuine ${\cal O}(\alpha)$ corrections to facilitate a comparison
to calculations made by other groups in the future.
Table~\ref{tab:relOalpha} shows the relative ${\cal O}(\alpha)$ corrections,
i.e.\
without higher-order ISR improvements and without QCD corrections,
both for the DPA of
{\sc RacoonWW} and for the full $\Pep\Pem\to4f$ calculation
for the three final states $\nu_\tau\tau^+\mu^-\bar\nu_\mu$,
$\Pu\bar\Pd\mu^-\bar\nu_\mu$, and
$\Pu\bar\Pd\Ps\bar\Pc$.
\begin{table}
$\begin{array}{c|cc|cc|cc}
\Pep\Pem\to{} &
\multicolumn{2}{c|}{\nu_\tau\tau^+\mu^-\bar\nu_\mu} &
\multicolumn{2}{c|}{\Pu\bar\Pd\mu^-\bar\nu_\mu} &
\multicolumn{2}{c}{\Pu\bar\Pd\Ps\bar\Pc}
\\
\hline
\sqrt{s}/\mathrm{GeV} & \de_{\DPA}[\%] & \de_{\eefourf}[\%] &
\de_{\DPA}[\%] & \de_{\eefourf}[\%] & \de_{\DPA}[\%] & \de_{\eefourf}[\%]
\\
\hline
161  & -31.30(3) & -29.12(3) & -31.30(3) & -29.15(3) & -31.27(3) & -29.21(3)
\\
170  & -21.67(2) & -21.10(2) & -21.75(2) & -21.16(2) & -21.81(2) & -21.23(2)
\\
189  & -11.74(2) & -11.57(2) & -11.84(2) & -11.65(2) & -11.94(2) & -11.74(2)
\\
200  & -9.15(2)  & -9.01(2)  & -9.25(2)  & -9.09(2)  & -9.33(2)  & -9.17(2)
\\
500  & +3.55(3)  & +3.30(4)  & +3.50(3)  & +3.21(4)  & +3.37(3)  & +3.08(4)
\\
1000 & +7.08(5)  & +6.43(6)  & +7.16(5)  & +6.43(7)  & +6.89(5)  & +6.15(7)
\end{array}$
\caption{Genuine relative ${\cal O}(\alpha)$ corrections, i.e.\
without higher-order ISR improvements and QCD corrections.}
\label{tab:relOalpha}
\end{table}

\subsection{Remaining theoretical uncertainties}

\begin{sloppypar}
We have reduced the TU for the charged-current processes
$\Pep\Pem\to\nu_\tau\tau^+\mu^-\bar\nu_\mu$,
$\Pu\bar\Pd\mu^-\bar\nu_\mu$, 
$\Pu\bar\Pd\Ps\bar\Pc$,
in particular in the threshold region of W-pair production, 
considerably by calculating the full $\Oa$ corrections.  ISR beyond
$\Oa$ is included via structure functions in leading-logarithmic
accuracy.
For energies below $\sim500\GeV$, the remaining uncertainties 
resulting from missing electroweak corrections
are then dominated by the
next-to-leading logarithmic electromagnetic corrections of order
$(\al/\pi)^2\log(\Me^2/s)$ which can be estimated to contribute
$\lsim0.1\%$. Near the W-pair-production threshold, higher-order
effects of the Coulomb singularity are still missing.
These are estimated to $\sim 0.2\%$
\cite{Fadin:1995fp,Bardin:1993mc}. Thus, we estimate the theoretical
uncertainty due to unknown electroweak higher-order effects
in the present calculation to be a few $0.1\%$ from the
threshold region to about $\sim500\GeV$. At higher energies leading
and subleading electroweak high-energy logarithms, such as Sudakov
logarithms, beyond one loop have to be taken into account in addition
to match this accuracy.

For a thorough estimate of the total theoretical uncertainty an
inclusion of QCD effects is indispensable for the processes involving 
final-state quarks. In order to reach a precision of the order of a
few 0.1\% there, it is certainly necessary to improve the treatment
of ${\cal O}(\alpha_{\mathrm{s}})$ corrections (and beyond), including
a proper matching with parton showers. Bose--Einstein and 
colour interconnection effects may also play an important role.
\looseness -1
\end{sloppypar}

\section{Summary}
\label{se:sum}

We have presented results on total cross sections for the
charged-current four-fermion production processes
$\Pep\Pem\to\nu_\tau\tau^+\mu^-\bar\nu_\mu$,
$\Pu\bar\Pd\mu^-\bar\nu_\mu$, 
$\Pu\bar\Pd\Ps\bar\Pc$ which, for the first time, include
the complete electroweak ${\cal O}(\alpha)$ corrections.
The calculation is consistently performed using complex gauge-boson
masses, supplemented by complex couplings to restore gauge invariance.
The evaluation of the occurring one-loop tensor integrals,
which include 5- and 6-point functions, required new techniques.
Moreover, we have developed algorithms to reduce the large number of
different spinor chains to a set of very few standard structures.

A comparison with the predictions for the total cross section without
cuts based on the DPA provided by the generator {\sc RacoonWW} reveals
corrections beyond DPA of $\lsim 0.3\%\;(0.6\%)$ for CM energies ranging
from $\sqrt{s}\sim200\GeV$ ($170\GeV$) to $500\GeV$,
consistent with previous estimates on the intrinsic DPA uncertainty.
The difference to the DPA increases to $0.7{-}1.6\%$ for
$\sqrt{s}\sim1{-}2\TeV$.  At threshold, where
predictions had to be based on an IBA at LEP2, the full ${\cal
  O}(\alpha)$ calculation corrects the IBA by about $2\%$, also
consistent with a previous error estimate.  The full ${\cal
  O}(\alpha)$ calculation, improved by higher-order effects from ISR,
reduces the remaining TU due to unknown electroweak higher-order effects
to a few $0.1\%$ for scattering energies from
the threshold region up to $\sim500\GeV$; above this energy leading
high-energy logarithms, such as Sudakov logarithms, beyond one loop
have to be taken into account to match this accuracy.
At this level of accuracy, also improvements in the treatment of QCD
corrections to semileptonic and hadronic $\Pep\Pem\to4f$ processes
will be necessary in the future.

Results for differential cross sections as well as details of the
calculation will be presented elsewhere.

\section*{Acknowledgements}

This work was supported in part by the Swiss National Science
Foundation, by the Swiss Bundesamt f\"ur Bildung und Wissenschaft, and
by the European Union under contract HPRN-CT-2000-00149.

\end{document}